# Geographical and Disciplinary Coverage of Open Access Journals: OpenAlex, Scopus and WoS

Abdelghani Maddi
abdelghani.maddi@cnrs.fr
GEMASS – CNRS – Sorbonne Université, 59/61 rue Pouchet 75017 Paris, France.
ORCID: http://orcid.org/0000-0001-9268-8022

Marion Maisonobe
marion.maisonobe@cnrs.fr
ORCID: https://orcid.org/0000-0002-2968-9038
Laboratoire Géographie-cités, CNRS, Université Paris 1, Université Paris Cité, EHESS, Aubervilliers, France

Chérifa Boukacem-Zeghmouri
cherifa.boukacem-zeghmouri@univ-lyon1.fr
Université Claude Bernard Lyon-1, Villeurbanne, France.
ORCID: https://orcid.org/0000-0002-0201-6159

## Abstract

This study aims to compare the geographical and disciplinary coverage of OA journals in three databases: OpenAlex, Scopus and the WoS. We used the ROAD database, managed by the ISSN International Centre, as a reference database which indexes 62,701 OA active resources (as of May 2024). Among the 62,701 active resources indexed in the ROAD database, the Web of Science indexes 6,157 journals, while Scopus indexes 7,351, and OpenAlex indexes 34,217. A striking observation is the presence of 25,658 OA journals exclusively in OpenAlex, whereas only 182 journals are exclusively present in WoS and 373 in Scopus.

The geographical analysis focusses on two levels: continents and countries. As for disciplinary comparison, we use the ten disciplinary levels of the ROAD database. Moreover, our findings reveal a striking similarity in OA journal coverage between WoS and Scopus. However, while OpenAlex offers better inclusivity and indexing, it is not without biases. WoS and Scopus predictably favor journals from Europe, North America and Oceania. Although OpenAlex presents a much more balanced indexing, certain regions and countries remain relatively underrepresented. Typically, Africa is proportionally as under-represented in OpenAlex as it is in WoS, and some emerging countries are proportionally less represented in OpenAlex than in WoS and Scopus.

These results underscore a marked similarity in OA journal indexing between WoS and Scopus, while OpenAlex aligns more closely with the distribution observed in the ROAD database, although it also exhibits some representational biases.

## Keywords
Open Access, OpenAlex, bibliographic databases, journals coverage, bibliometrics.

## Data availability
Data will be available in Zenodo

## Conflict of interest
The authors have no relevant financial or non-financial interests to disclose.

## Acknowledgments

Authors would like ISSN International Centre for providing the ROAD data. We would like to express our heartfelt gratitude to **Dr. Gaëlle BEQUET**, Director of the ISSN International Centre, for reviewing the initial draft of this paper. Her insightful comments and constructive feedback have been invaluable in enhancing the quality and depth of our work. We sincerely appreciate her time and effort, which have significantly contributed to the improvement of this article.

## Introduction

The academic publishing landscape is experiencing a seismic shift as more researchers, institutions, and funding bodies embrace open access (OA) publishing models (Lewis, 2017). This transition is driven by a multifaceted set of factors, including policy mandates, institutional initiatives, and changing attitudes towards scholarly communication. Governments, research funders, and academic institutions worldwide are increasingly recognizing the benefits of making research outputs freely accessible to all, without barriers such as subscription fees or paywalls (Tennant et al., 2016).

In response to these incentives, there has been a proliferation of OA journals and platforms, facilitated by various funding models (Pölönen et al., 2021; Ross-Hellauer et al., 2018). One notable development is the emergence of the Diamond model, characterized by journals that are both OA and free of publication charges for authors (Fu et al., 2024; Fuchs & Sandoval, 2013). Traditional bibliographic databases like Web of Science (WoS) and Scopus were developed on the basis of the subscription-based journals, and their coverage did not fully take into account the development of the growing prevalence of OA publishing (Severin et al., 2020). While these databases have made efforts to incorporate OA content, they often lag behind in terms of coverage and inclusivity (Basson et al., 2022). This discrepancy is particularly evident when comparing their journal coverage to that of databases like the Directory of OA Scholarly Resources (ROAD), which exclusively indexes OA journals.

Enter OpenAlex, a new player in the field of bibliographic databases, heralded at least in France and some European countries, as a potential game-changer in the realm of OA publishing (Cloutier & Dacos, 2023). OpenAlex agenda is to address the limitations of traditional databases by providing a more comprehensive and inclusive index of scholarly journals (Priem et al., 2022). By leveraging advanced data harvesting techniques and partnerships with academic institutions, OA repositories and publishers, OpenAlex aims to offer a broader and more diverse representation of scholarly outputs from around the world (Alperin et al., 2024). The current key question surrounding OpenAlex is whether it can effectively address the historical biases and limitations of traditional databases. While OpenAlex holds promise as a potential solution to these longstanding issues, its effectiveness remains to be fully evaluated. Ongoing research and analysis will be essential to assess the impact of OpenAlex on the visibility, accessibility, and diversity of scholarly publications. As the academic publishing landscape continues to evolve, OpenAlex represents a significant development in the ongoing quest for a more open, inclusive, and equitable scholarly communication ecosystem.

Aiming to contribute to this topic, our research questions are:

**RQ1:** *Will OpenAlex provide better coverage of OA journals, including those from under-represented regions?*

**RQ2:** *Can it overcome the challenges of disciplinary biases and ensure a more equitable representation of research outputs across diverse fields?*

In this study, our objective is to conduct a comparative analysis of the geographical and disciplinary coverage of OA journals across three prominent databases: Web of Science (WoS), Scopus, and OpenAlex. Through this approach, we aim to shed light on the differences and nuances in OA journal coverage across the three databases. For this purpose, we employ the ROAD database as a coverage reference as it provides a comprehensive catalog of 62,701 indexed OA journals. Our analysis focusses on two key dimensions: geographical representation and disciplinary diversity. Geographically, we examine coverage across continents and countries. Additionally, we conduct a disciplinary comparison using the ten disciplinary levels outlined in the ROAD database.

The article is structured as follows: we start with a review of the literature on recent papers comparing the three databases. Then, we present the data collected and method used aligned on the study objectives. The results encompass an analysis of the overall coverage of OA journals and a comparison of their geographical and disciplinary structure, including an examination at the country and continent levels. Finally, we provide a discussion of the findings and their implications.

# Literature review

While there is a growing number of studies using OpenAlex as a data source, few have focused specifically on OpenAlex's OA coverage and limitations discussed in this paper. This is why our literature review focuses solely on comparative studies involving OpenAlex alongside other sources, a choice aligned with our research questions.

A recent study by (Alperin et al., 2024) observed a growing trend in using OpenAlex as a data source, but noted a lack of studies focusing on OpenAlex itself and its limitations. They compared OpenAlex and Scopus data, finding more publications in OpenAlex, particularly from regions and languages under-represented in Scopus. However, they identified areas for improvement in metadata accuracy (e.g., affiliations, document types, open access status) and completeness in OpenAlex. Another study by (Culbert et al., 2024) explored OpenAlex as a promising open source of scholarly metadata, comparing it with Web of Science and Scopus (whose metadata are not error-free either). They assessed reference and metadata coverage, demonstrating OpenAlex's comparability in reference numbers but mixed results in other metadata. Their study highlighted the importance of addressing data and metadata trustworthiness in rapidly evolving sources like OpenAlex.

(Jiao et al., 2023) investigated the indexing of data papers in scholarly databases to understand how research data is published and reused. They examined 18 data journals across WoS, Scopus, Dimensions, and OpenAlex to evaluate coverage and document type information consistency. Their findings revealed highly inconsistent coverage of data papers and their document types across databases, posing challenges for quantitative analysis. (Jiao et al., 2023) showed that, while newer databases like Dimensions and OpenAlex cover all exclusively data journals, they classify data papers as regular research articles, making their retrieval challenging. In contrast, although Scopus and WoS cover fewer data journals, they distinguish data papers with a "Data paper" document type. However, inconsistencies persist, indicating a need for improved communication to enhance database quality.

(Ortega & Delgado-Quirós, 2023) analyzed retractions and withdrawals in scholarly databases, highlighting differences between traditional citation indexes like the WoS and newer hybrid databases like OpenAlex. Their findings underscored the impact of database selection on coverage of retractions and withdrawals. (Ortega & Delgado-Quirós, 2023) highlighted that the differences primarily stem from how withdrawals are indexed by newer hybrid databases like Dimensions, OpenAlex, Scilit, and The Lens. Excluding withdrawal data, OpenAlex and The Lens collect the most retractions, while Scilit, Scopus, and Dimensions include the highest number of retracted articles. This suggests a distinction between traditional citation indexes such as WoS, PubMed, and Scopus, which are journal-based and do not index withdrawals, and newer hybrid databases relying on external sources like Crossref and Microsoft Academic. Since September 2023, Crossref, the leading DOI registration agency, acquired the RetractionWatch database, making it freely accessible (Hauschke & Nazarovets, 2024). Consequently, OpenAlex now directly incorporates RetractionWatch data to enrich its retraction field (see: https://docs.openalex.org/api-entities/works/work-object). More recently, (Delgado-Quirós & Ortega, 2024) aimed to compare metadata completeness across academic databases. They found that third-party databases like OpenAlex had higher metadata quality and completeness compared to academic search engines like Google Scholar. Their study emphasized the need for reliable descriptive data retrieval, especially in third-party databases.

Moreover, other studies point out that the metadata quality of OpenAlex has significant room for improvement to be usable in bibliometric studies. For example, (Zhang et al., 2024) investigated missing institutional information in journal article metadata in OpenAlex. They identified significant gaps, particularly in early years and social sciences and humanities. Their study emphasized the importance of data quality improvements in open resources like OpenAlex. Similarly, (Bordignon, 2024) discussed the growing adoption of OpenAlex, citing institutions' decisions to transition from proprietary bibliometric products, notably the decision of Sorbonne Université in France to unsubscribe from Web of Science[1], and CNRS (French national research organism) unsubscribing from Scopus[2] but keeping WoS subscription. She highlighted the importance of assessing the relevance of OpenAlex for bibliometric analysis, presenting tests to evaluate its effectiveness at an institutional level. (Bordignon, 2024) came out with similar conclusions to (Zhang et al., 2024) regarding the quality of institutional metadata, based on a case study of publications from École des Ponts (a French engineering school).

More recently, (Céspedes et al., 2024) assessed the linguistic coverage of OpenAlex and the accuracy of its metadata compared to Web of Science (WoS). Through an in-depth manual validation of 6,836 articles, the study found that OpenAlex offers a more balanced representation of non-English languages than WoS. However, the language metadata was not always accurate, leading to an overestimation of English and an underestimation of other languages. This research underlines the need for infrastructural improvements to ensure accurate metadata, despite OpenAlex's potential for comprehensive linguistic analysis in scholarly publishing. Another study, (Alonso-Alvarez & van Eck, 2024) examined the coverage and metadata availability of African publications in OpenAlex, comparing it with Scopus, WoS, and African Journals Online (AJOL). Their findings revealed that OpenAlex offers the most extensive coverage of African-based publications, but still lags in providing detailed metadata, particularly regarding affiliations, references, and funder information. Interestingly, metadata completeness was found to be better for publications indexed in both OpenAlex and the proprietary databases, highlighting areas for improvement in OpenAlex to better serve research from the Global South.

# Data

Data collected for the study was primarily sourced from the Directory of OA Scholarly Resources (ROAD) (https://www.issn.org/services/online-services/road-the-directory-of-open-access-scholarly-resources/), kindly provided to us by the ISSN International Centre (https://www.issn.org/) in XML format. We extracted information for each of the 62,701 indexed sources (ISSN), including country (across 163 countries) and disciplines (10 levels). It is noteworthy that some journals lacked information regarding their discipline (3.6 % - 2263 journals out of 62,701).

Launched in late 2013, ROAD offers free access to a subset of bibliographic records from the ISSN Portal, describing scholarly resources available in OA identified by an ISSN. These resources include journals, monograph series, conference proceedings, open archives, institutional repositories, and research blogs. Metadata for these records, created by the ISSN network comprising 93 national centres and the ISSN International Centre, are enriched with data from indexing services, directories (such as DOAJ, Latindex, The Keepers Registry), and performance indicators (Scopus). ROAD is aligned with UNESCO's efforts to promote OA to scholarly resources and complements the Global Open Access Portal (GOAP) (https://www.goap.info/) developed by UNESCO, which provides an overview of OA to scholarly information worldwide.

---

[1] See: https://www.sorbonne-universite.fr/actualites/sorbonne-universite-se-desabonne-du-web-science#:~:text=C'est%20pourquoi%20Sorbonne%20Universit%C3%A9,aux%20outils%20bibliom%C3%A9triques%20de%20Clarivate.

[2] https://www.cnrs.fr/en/update/cnrs-has-unsubscribed-scopus-publications-database#:~:text=The%20organisation%20has%20taken%20an,open%20solutions%20are%20sufficiently%20mature.

ROAD applies specific inclusion criteria to list resources in its directory. To be included, a resource needs several requirements, including being freely accessible without registration, providing a clear description of its OA policy and licensing terms, and presenting scholarly content across various fields. The resource must also have clear editorial responsibility, academic affiliation, publishing entity, and adhere to ethical guidelines and indexing standards.

For the purpose of this study, ROAD served as the gold standard for comparing the coverage in OA journals of OpenAlex, Scopus, and WoS. The data extraction was conducted in October 2023, yielding 253,200 identifiers (ISSN and/or EISSN) for 183,158 distinct journals from OpenAlex. We limited our analysis to active journals in OpenAlex with more than five publications: out of 183,158 journals in OpenAlex, 6,632 have five or fewer publications and were excluded from the analysis. As a result, we considered 176,526 journals from OpenAlex, 29,262 active journals from Scopus, and 23,189 from the WoS Core Collection (AHCI, SSCI, SCIE, and ESCI).

For income data by country, we used the R package tmap (Tennekes, 2018), which provides this information. Additionally, we enriched the metadata for 13 territories that were not listed in the 'World' dataset of tmap[3].

# Methods

The study has a dual objective: 1) to analyze the overall coverage of OA journals in the three databases and 2) to investigate the geographical and disciplinary distribution to assess the extent to which disciplines, countries, and regions are represented in each database.

For the analysis of overall coverage, we employed an UpSet graph to visualize the coverage and intersections among the three databases. This graphical representation allows for a comprehensive examination of the shared and unique journals indexed by each database.

Regarding the distribution analysis, an indicator (the $Coverage\ Index_{ij}$) was calculated to assess the representation of various entities (such as countries, continents, or disciplines) in each of the three databases (WoS, Scopus, and OpenAlex), relative to their representation in the entire ROAD database.

$$Coverage\ Index_{ij} = \frac{OA\ journals\ indexed_{ij} / All\ OA\ journals\ indexed_j}{OA\ journals\ indexed_{i\ ROAD} / All\ OA\ journals\ indexed_{ROAD}}$$

Where $i$ the entity (continent, discipline, etc.), $j$ the database (WoS, Scopus and Scopus).

Specifically, this indicator compares the proportion of a given entity (e.g., a country) within the journals indexed by a particular database (e.g., OpenAlex) to the proportion of that entity within the entire ROAD database. For example, to calculate the indicator for Italy in OpenAlex, the proportion of Italy within the journals indexed by OpenAlex (1.6%) is divided by the proportion of Italy within the entire ROAD database (7%). The neutral value of this indicator is 1. Therefore, if a given country has an indicator of 1.30 in WoS, it would indicate that it is overrepresented by 30% in this database compared to the global structure of OA journals.

---

[3] Namely the territories with the following ISO3 codes: BHR (Bahrain), BRB (Barbados), MLT (Malta), MUS (Mauritius), SGP (Singapore), SYC (Seychelles), GLP (Guadeloupe), GUF (French Guiana), REU (Réunion), MTQ (Martinique), GUM (Guam), MAC (Macau), HKG (Hong Kong).

This method allows for a refined assessment of the representation biases within each database, highlighting any discrepancies in the geographical and disciplinary distribution of OA journals across the three platforms.

## Results

As shown in Figure 1, among the journals indexed in the ROAD database, the WoS indexes 6,157 journals, while Scopus indexes 7,351, and OpenAlex indexes 34,217. A striking observation is the presence of 25,658 OA journals exclusively in OpenAlex, whereas only 182 journals are exclusively present in WoS and 373 in Scopus. Additionally, 4,104 OA journals are simultaneously indexed in all three databases, while 145 journals are indexed simultaneously in WoS and Scopus but not in OpenAlex.

**Figure 1: Comparison of coverage of OA journals in Web of Science, Scopus, and OpenAlex**

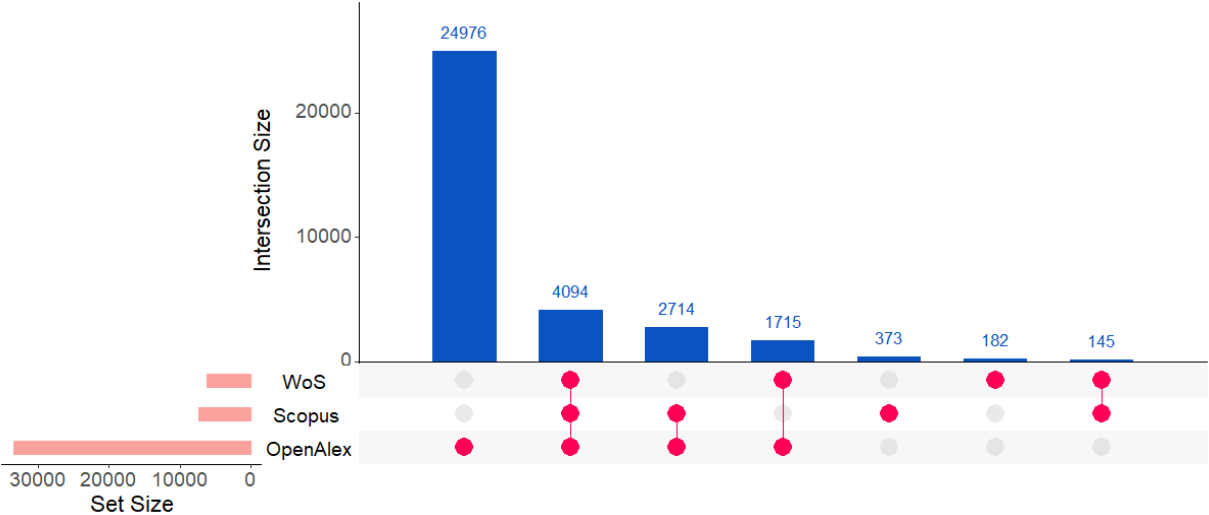

These results underscore the important role of OpenAlex in expanding the coverage of OA journals, with a significant proportion of journals exclusive to this database. However, it is crucial to acknowledge that some journals may publish a small number of publications, as evidenced by a recent study focusing on Diamond journals (Bosman et al., 2021).

**Figure 2: Coverage index of databases by continent (neutral value = 1)**

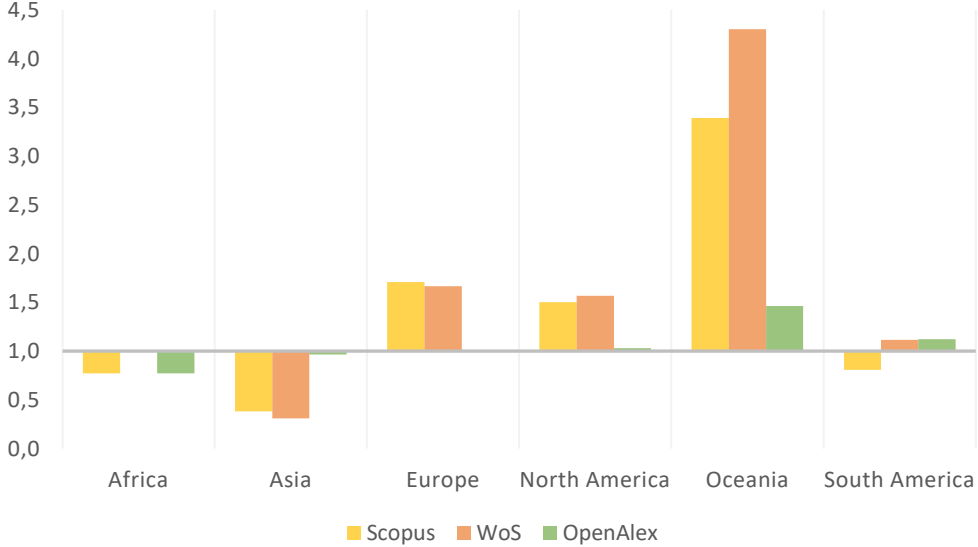

The examination of continent representation in Scopus, the WoS, and OpenAlex reveals distinct trends, highlighting disparities in journal coverage on a global scale. As represented in Figure 2, all databases exhibit significant differences in continent representation. While Scopus and WoS generally show similar patterns, OpenAlex stands out with different representation trends. Scopus and WoS, despite minor variations in their continental indices, demonstrate a tendency towards overrepresentation of Oceania, North America and Europe, along with a relative under-representation of Africa and Asia. These databases seem to favor journal coverage from hegemonic regions, raising questions about equity and inclusivity in their indexing practices.

Conversely, OpenAlex presents more diverse representation patterns, with a trend towards better geographical equity. Although some disparities persist, such as the underrepresentation of Africa and Asia, OpenAlex appears to offer more inclusive coverage of journals from various regions worldwide.

The high values of Oceania in Scopus (3.39) and WoS (4.30) indicate that the region's presence in these databases is more than three and four times higher than the global average, respectively. This high representation is primarily driven by the significant contribution of Australia and New Zealand, both of which have robust research output and strong academic infrastructures. Their focus on publishing in high-impact, English-language journals aligns well with the indexing criteria of Scopus and WoS, leading to their disproportionate visibility. This imbalance reflects a linguistic and systemic bias, where English-speaking countries with well-established research ecosystems are more prominently featured, thereby inflating the overall regional values in these international bibliometric databases. Australia, in particular, largely benefited from the Regional Expansion of the Web of Science in 2006-2008. However, this is less the case in OpenAlex (1.46), which offers a more balanced representation of global research by including a broader range of OA sources, including those in multiple languages and from less prominent regions. Interestingly, Scopus and OpenAlex exhibit a common over-representation pattern regarding the coverage of South American OA journals whereas OA journals from this continent are under-represented in the WoS.

**Figure 3: Coverage index by database for different income groups**

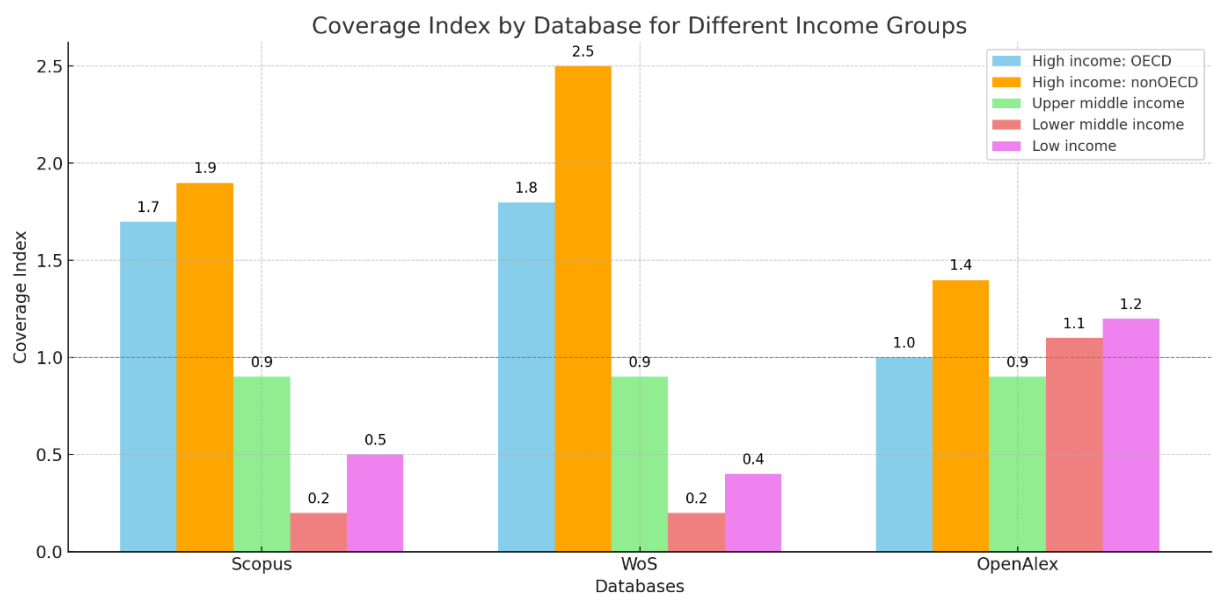

The comparison of economic income (figure 3) groups across Scopus, WoS, and OpenAlex databases reveals also notable disparities in representation. High-income economies, whether OECD or non-OECD, are consistently over-represented in Scopus and WoS, reflecting historical biases. Unlike WoS and Scopus, OpenAlex operates under a fundamentally different paradigm, prioritizing an open and inclusive indexing strategy. While WoS and Scopus often rely on selective curation processes that may favor established publishers and higher-income regions, OpenAlex harvesting strategy embraces a broader, more decentralized approach. This allows for a wider representation of research outputs across various economic contexts, not necessarily through targeted efforts but rather through a strategic commitment to openness and accessibility that contrasts with the more traditional, restrictive models of other platforms. Middle-income economies show varying levels of representation, with indicators fluctuating across databases.

Lower middle-income economies are particularly underrepresented in Scopus and WoS, highlighting systemic biases within these databases. OpenAlex presents a more equitable representation across middle-income categories. Low-income economies face consistent under-representation in Scopus and WoS, indicative of broader challenges in access to scholarly resources. While OpenAlex offers a more inclusive platform, potential overrepresentation raises questions about indexing criteria and inclusivity.

Analyzing the representation of disciplines across Scopus, WoS, and OpenAlex databases provides insights into the distribution of scholarly knowledge and potential biases within each database. Figure 4 comparing disciplinary representation across these databases reveals notable variations in coverage, reflecting broader trends in academic publishing and bibliographic indexing practices.

Scopus and WoS demonstrate consistent biases towards certain disciplines, particularly within STEM fields such as Physics and Natural Sciences. These databases prioritize journals with high impact factors, often favoring quantitative research outputs and publications from well-established and high-ranked academic institutions. Consequently, disciplines within the humanities and social sciences may be under-represented in Scopus and WoS, reflecting historical publishing and citation biases. Conversely, OpenAlex presents a more inclusive approach to indexing, aiming to encompass a diverse range of scholarly outputs across disciplines. While OpenAlex exhibits slightly lower representation in some disciplines compared to Scopus and WoS, such as Applied Sciences, Medicine, and Technology, it offers a more balanced representation overall. This suggests that OpenAlex's indexing practices may be more reflective of the diverse disciplinary landscape and less influenced by traditional biases prevalent in academic publishing. Notably, academic subjects such as "Social sciences" demonstrate higher representation in OpenAlex compared to Scopus and WoS, highlighting potential differences in indexing criteria and inclusivity across databases. This variation underscores the crucial importance of considering multiple databases to ensure comprehensive coverage across diverse disciplinary areas.

**Figure 4: Coverage index of databases by discipline (neutral value = 1)**

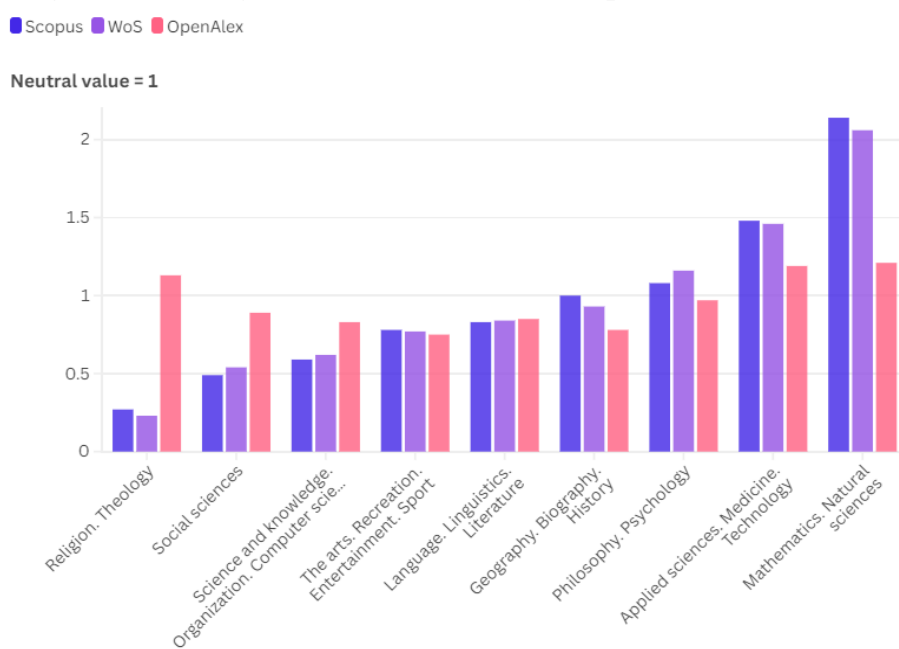

Figures 5a, 5b, and 5c highlights the differences in the coverage of open-access (OA) journals between the Web of Science (WoS), Scopus, and OpenAlex databases at the country level. The world maps represent the geographical distribution of the coverage indicator for each database (described in the Methods section), showcasing well-represented regions and countries as well as underrepresented ones. The circle packings show the share of ROAD journals covered by each database (color gradient) whereas the size of the circles depend on the absolute number of ROAD journal per country. The similarity of the WoS and Scopus compared to OpenAlex is really striking on these representations.

Upon examining Figure 5, it becomes evident that certain regions, such as Western Europe and North America, are well represented in all three databases, with high index values. However, significant disparities emerge for other parts of the world. For instance, some areas in Africa, Asia, and Latin America exhibit relatively low coverage indices in WoS and Scopus but better representation in OpenAlex. This difference can be attributed to OpenAlex's more inclusive indexing policy, which covers a broader range of OA journals from diverse regions around the world. In contrast, WoS and Scopus may exhibit geographic biases, often favoring journals published in specific countries or regions, and applying selection processes that emphasize certain quality standards, which can further limit the diversity of represented sources.

This figure allows us to observe to what extent OpenAlex provides a more inclusive and balanced indexing of OA journals compared to both WoS and Scopus, which show similar coverage indices and comparable geographic biases. At the same time, it is worth noting that each database has its own strengths and weaknesses, and the choice depends on the specific needs of each research endeavor.

Let us now focus on two outliers. In Europe, the case of France stands out as a notable exception. Despite its status as a developed country with a substantial scholarly output, French OA journals are consistently underrepresented across all three databases. This discrepancy raises questions about the systemic factors contributing to the exclusion of French journals from mainstream bibliographic databases. However, there is a possibility of French journals being disproportionately represented in the ROAD database, the reference database used for comparison. This potential over-representation in ROAD could skew perceptions of under-representation in Scopus, WoS, and OpenAlex. Further investigation into this discrepancy is necessary to understand the underlying factors contributing to the exclusion or under-representation of French OA journals.

Finally, the case of Indonesia also raises similar questions. Specifically, both OpenAlex and ROAD exhibit a notable pattern for this country. Indonesia emerges as the top contributor to the ROAD database (it is the biggest circle on the circle packing plots). One possible explanation is that, since 2019, the Indonesian government has mandated through legislation that universities must create OA journals (Priadi et al., 2020). This policy has significantly increased the number of Indonesian journals in ROAD, potentially leading to an apparent over-representation in this database. As for France, this scenario necessitates further investigation to understand the impact of such policies on the visibility and representation of Indonesian journals in bibliographic databases like OpenAlex.

**Figure 5: Coverage index of OA journals by database by country**

| Figure 5a. Coverage index by country in the WoS | Figure 5b. Coverage index by country in Scopus | Figure 5c. Coverage index by country in OpenAlex |
|---|---|---|
| 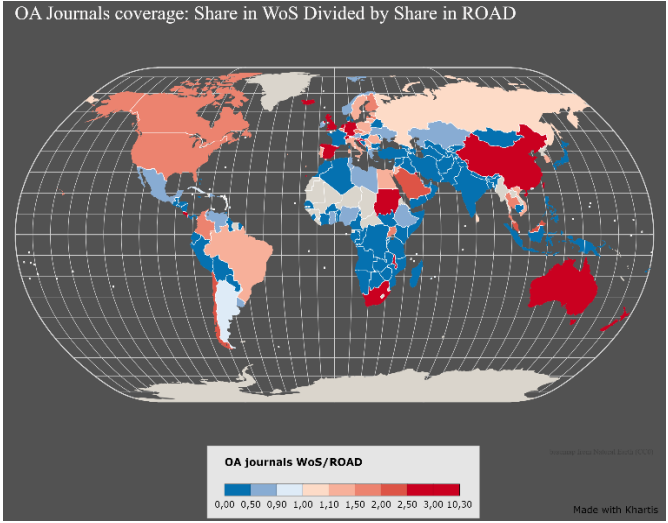 | 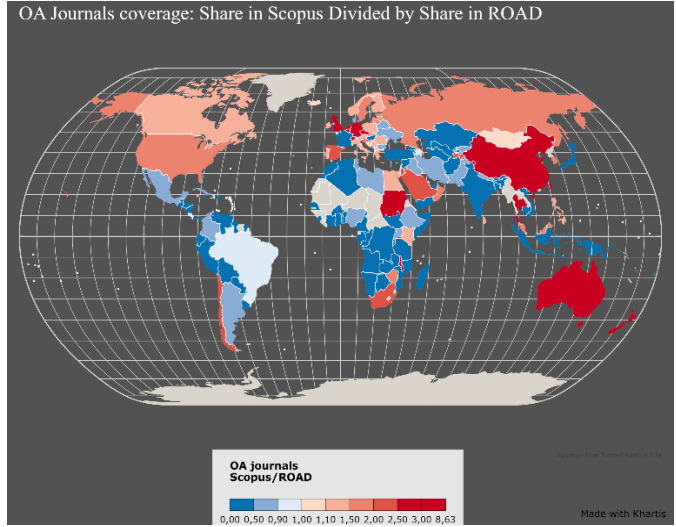 | 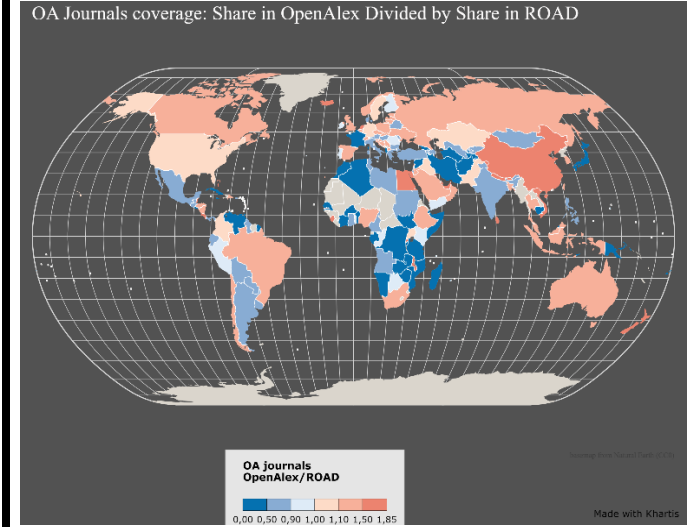 |
| 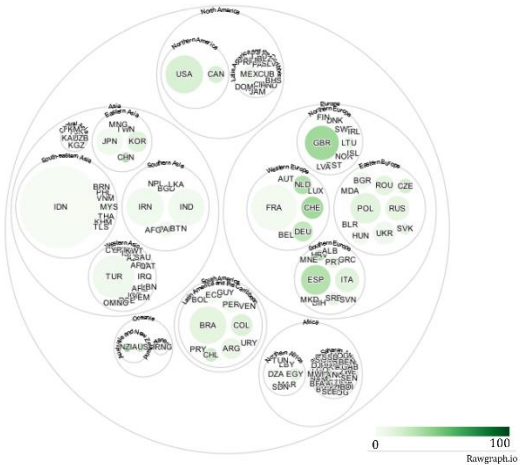 | 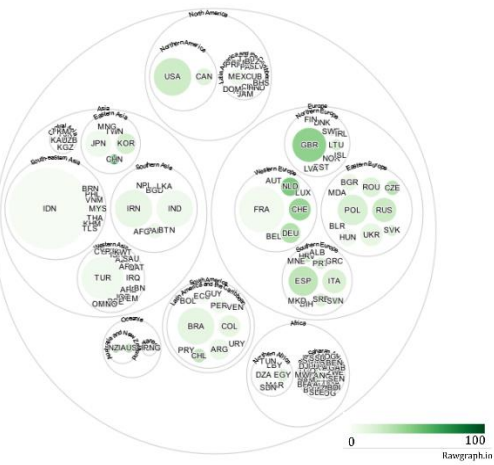 | 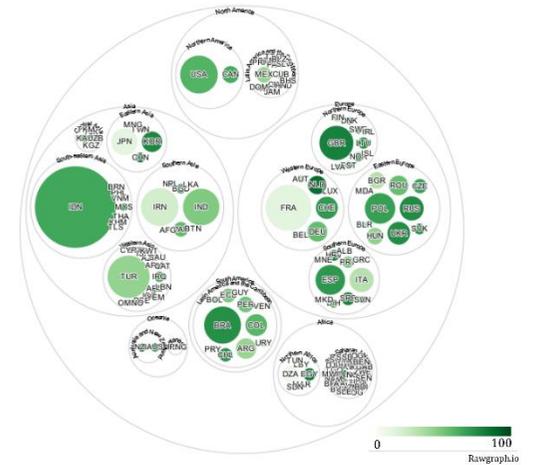 |

Share of ROAD journals included in the Web of Science (%). The circles' size is proportionnal to the absolute number of ROAD journals per country.

Share of ROAD journals included in Scopus (%). The circles' size is proportionnal to the absolute number of ROAD journals per country.

Share of ROAD journals included in Open Alex (%). The circles' size is proportionnal to the absolute number of ROAD journals per country.

# Discussion

This study presents a comprehensive assessment of the coverage of OA journals across three major scholarly databases: Web of Science (WoS), Scopus, and OpenAlex. The analysis reveals significant variations in representation across different countries, regions, development levels, continents, and income groups, highlighting the importance of database selection when it comes to scholarly research and outputs.

The examination of coverage indices highlights notable differences in the representation of countries and regions and disciplines across the three databases. While certain areas, such as Western Europe and North America, enjoy robust coverage across all platforms, disparities are evident in other regions. OpenAlex confirming its agenda exhibits a more inclusive approach, capturing a broader range of OA journals from diverse geographic locations compared to WoS and Scopus. This suggests that OpenAlex serves as a valuable resource for researchers seeking comprehensive coverage, especially for studies focusing on under-represented regions (as soon as the biases are known, i.e. the case of Indonesia).

Furthermore, the analysis points out disparities in coverage based on the development level of countries. Developing regions, particularly those in Africa and parts of Asia, tend to have lower coverage indices in WoS and Scopus compared to more developed regions. OpenAlex demonstrates a more equitable representation of journals from developing countries, offering researchers access to a wider array of scholarly content. Likewise, analysis based on income groups further highlights disparities in database coverage. Low and middle-income countries often experience lower representation in WoS and Scopus compared to high-income countries. OpenAlex demonstrates a more equitable distribution of journals across income groups.

Finally, our findings underscore the importance of database selection in shaping scholarly research and knowledge dissemination. OpenAlex emerges as a new and valuable source for researchers seeking a more comprehensive coverage of OA journals, particularly from underrepresented regions, continents, development levels, and income groups.

**Conclusion**

In conclusion, our study highlights the role of database selection in shaping the landscape of scholarly research and knowledge dissemination. The disparities in coverage across Web of Science (WoS), Scopus, and OpenAlex underscore the varying degrees of inclusion and representation of OA journals from different countries, regions, continents, development levels, and income groups. OpenAlex, with its broader and more inclusive approach, emerges as a valuable resource for researchers, particularly those focusing on underrepresented regions or countries with lower income levels. This platform mitigates some of the biases inherent in traditional databases like WoS and Scopus, offering a more equitable distribution of OA journals.

To better understand our results on France representation, a closer examination is needed to explore how national policies influence the visibility and inclusion of Indonesian journals and other underrepresented regions in global bibliographic databases like OpenAlex. Future investigations could reveal the impact of such policies on enhancing the global visibility of scholarly work from less-represented regions, further promoting an equitable academic ecosystem.